\documentclass[conference]{IEEEtran}
\IEEEoverridecommandlockouts
\usepackage{cite}
\usepackage{amsmath,amssymb,amsfonts}
\usepackage{algorithmic}
\usepackage{graphicx}
\usepackage{textcomp}
\usepackage{xcolor}
\pdfoutput=1
\def\BibTeX{{\rm B\kern-.05em{\sc i\kern-.025em b}\kern-.08em
    T\kern-.1667em\lower.7ex\hbox{E}\kern-.125emX}}
\begin{document}

\title{Value-Decomposition Networks based Distributed Interference Control in Multi-platoon Groupcast\\
}

\author{\IEEEauthorblockN{ Xiongfeng Guo, Tianhao Wu and Lin Zhang}
\IEEEauthorblockA{\textit{School of Information and Communication Engineering} \\
\textit{Beijing University of Posts and Telecommunications}\\
Beijing, China \\
Email: \{guoxiongfeng, wu.tianhao, zhanglin\}@bupt.edu.cn}
}

\maketitle

\begin{abstract}
Platooning is considered one of the most representative 5G use cases. Due to the small spacing within the platoon, the platoon needs more reliable transmission to guarantee driving safety while improving fuel and driving efficiency. However, efficient resource allocation between platoons has been a challenge, especially considering that the channel and power selected by each platoon will affect other platoons. Therefore, platoons need to coordinate with each other to ensure the groupcast quality of each platoon. To solve these challenges, we model the multi-platoon resource selection problem as Markov games and then propose a distributed resource allocation algorithm based on Value-Decomposition Networks. Our scheme utilizes the historical data of each platoon for centralized training. In distributed execution, agents only need their local observations to make decisions. At the same time, we decrease the training burden by sharing the neural network parameters. Simulation results show that the proposed algorithm has excellent convergence. Compared with another multi-agent algorithm (MARL) and random algorithm, our proposed solution can dramatically reduce the probability of platoon groupcast failure and improve the quality of platoon groupcast.

\end{abstract}


\section{Introduction}
As a 5G new application scenario and Intelligent transportation system (ITS) service, the platoon has received more and more attention \cite{b1}. With very small headways, vehicles in platoons can achieve considerable fuel consumption gains \cite{b2}. To reap the benefits of platooning, platoon members must have enough awareness of the movement status of the platoon, assisted by the 5G vehicle-to-everything (V2X) communication technologies. However, in a multi-platoon scenario, the resource selection strategies, transmission range, and power of each platoon interact with each other, forming a complicated communication interference environment \cite{b3}. Improper radio resource allocation will cause interference to platoons using the same radio resources. Therefore, it is necessary to study the groupcast communication resource allocation algorithm for platoons to ensure the reliability of the groupcast in different platoons.

In recent years, many resource allocation schemes have been proposed, which can be divided into centralized and distributed schemes. In the centralized algorithm \cite{b3} \cite{b4}, the base station collects real-time information of different platoons, and then allocates resources to varying platoons according to algorithms such as branch and bound or graph theory to achieve system optimization. However, the high dynamics of the vehicle networking channel environment and the randomness of the V2V payload hinder the application of centralized algorithms in large-scale scenarios, which are reflected explicitly reflected in the complexity of the algorithm and the delay in information collection.

In the distributed algorithm, there is no need for the central controller to perform global scheduling. It can avoid wasting radio resources by signaling interaction, massive computing burden, and scalability limitations. Reinforcement learning (RL) is considered as a useful tool to solve the distributed allocation of 5G resources \cite{b5}. In RL, agents learn strategies to maximize system efficiency by observing changes in the environment. Recently, many intelligent resource management schemes using reinforcement learning have been proposed. In \cite{b6}, the author proposed a Q-learning based resource allocation algorithm. In \cite{b7}, a distributed resource allocation algorithm based on deep reinforcement learning was offered, assuming that the agents make asynchronous decisions. In \cite{b8}, a fingerprint-based deep Q-network method is introduced, which enabled agents to make decisions simultaneously. The above previous works have adopted specific processes such as asynchronous updates and fingerprints to reduce non-stationary in a multi-agent environment.

Unlike the above works, we propose a distributed resource allocation framework based on Value-Decomposition Networks (VDN). By integrating the value function of each agent, a joint action-value function is obtained, which is used for back-propagating by different individual deep neural networks. VDN guarantees that the agents cooperate to improve system efficiency. When deployed, the agent only needs its local observations to make optimal decisions, without the need for other signaling exchanges. The main contributions of this article can be summarized as
\begin{figure}[htbp]
    \centerline{\includegraphics[scale=0.4]{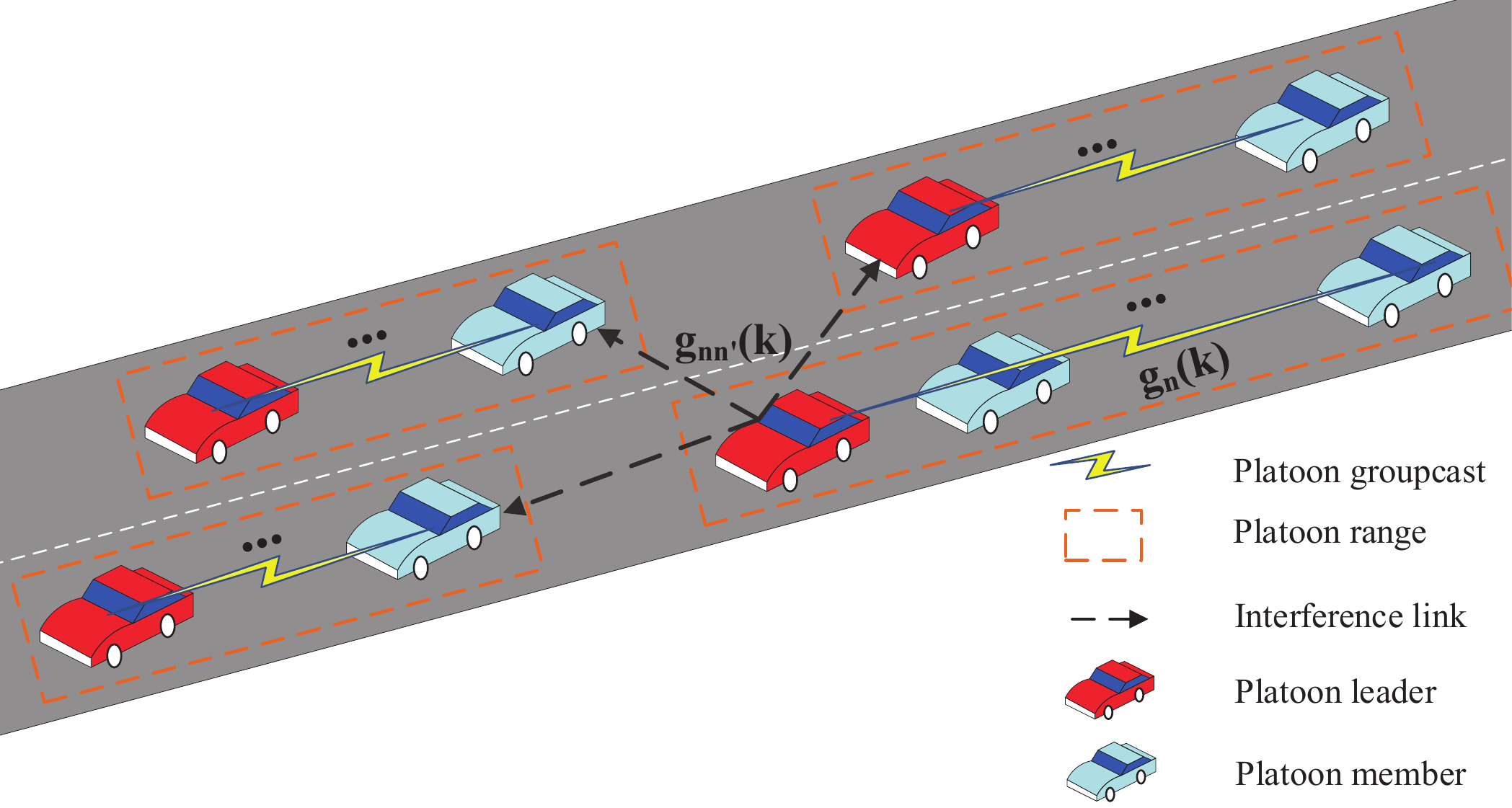}}
    \caption{The multi-platoon groupcast scenario.}
    \label{fig1}
\end{figure}
\begin{itemize}
    \item Considering the reliability requirements of the platoon groupcast, the hyperparameter $\tau$ is introduced to encourage the platoon to complete the transmission of the payload as soon as possible.
    \item The problem of collaborative interference control for multi-platoon in highway scenario is modeled as Markov games. We assume that each agent can only obtain its local observations, which cannot represent the entire environment. It can more accurately represent changes in the environment. 
    \item Designing spectrum access and power selection scheme based on VDN. Agents can coordinate their resource selection strategies through centralized training. During distributed execution, agents only need their observations to make global optimization decisions.
    \item The recurrent neural network (RNN) is used to integrate the agent's past actions and outputs to overcome the instability caused by local observability. And we let different agents share the neural network to speed up the training and convergence.
\end{itemize}

The rest of this paper can be organized as follows. In section \uppercase\expandafter{\romannumeral2}, we introduce the system model of multi-platoon firstly, then formulate optimization problems. In Section \uppercase\expandafter{\romannumeral3}, we model the resource allocation problem as a partially observable Markov game and propose a VDN based algorithm to solve this problem. Simulation results and analysis are presented in Section \uppercase\expandafter{\romannumeral4} and conclusions are drawn in Section \uppercase\expandafter{\romannumeral5}.

\section{System Model and Problem Formulation}
\subsection{System Model}
As shown in Fig.~\ref{fig1}, we consider a multi-platoon groupcast scenario. The set of platoons is denoted by $\mathcal{N}=\{1, \cdots, N\}$. Each platoon contains the platoon leader (PL) $V_{0}$ and the corresponding following vehicle $\left\{V_{1}, V_{2}, \cdots, V_{M}\right\}$, that is, the platoon members (PMs). To ensure the safety of the platoon, the PL must periodically groupcast information to the PMs. The dedicated spectrum of the platoon groupcast can be divided into multi-subchannel, denoted by $\mathcal{K}=\{1, \cdots, K\}$. $\theta^{k}=1$ represents that the PL chooses channel $k$ to groupcast the message, and $\theta^{k}=0$ otherwise. Assume that the transmit power selected by PL is $P_t$, and the receive power of the corresponding PMs can be expressed as $P_r=g_{i,j} P_t$, where $0\leq P_t \leq P_{max}$, $g_{i, j}$ represents the channel gain between the sender and receiver. We use a free-space propagation model and Rayleigh attenuation to model the channel gain. The channel gain can be expressed as
\begin{equation}
    g_{i,j}=\alpha_{i,j} h_{i,j}
\end{equation}
where $\alpha_{i,j}$ is the channel coefficient including path propagation fading, shadow fading and antenna gain, and its dB form expression is $\alpha_{i,j}^{dB}=G-PL(d_{i,j})-SH$. $h_{i,j}$ is a small-scale Rayleigh attenuation coefficient. In each communication link, $h_{i,j}$ conforms to an independent exponential distribution with a mean value of $1$. The signal-to-interference-plus-noise ratio (SINR) of receiver $j$ of the platoon $i$ in channel $k$ can be expressed as:
\begin{equation}
    \gamma_{i,j}^{k}=\frac{\theta_{i}^{k} P_{i} g_{i, j}}{\sigma^{2}+\sum_{i^{\prime}=1,i^{\prime} \neq i}^{N} \theta_{i^{\prime}}^{k} P_{i^{\prime}} g_{i^{\prime}, j}}
    \end{equation}
where $P_{i}$ represents the transmission power of the $i^{th}$ PL. $\sigma^{2}$ represents the noise power, and the other parts of the denominator represents interference from other PLs to $i^{th}$ platoon. We assume that each PL only selects one subchannel, and $\sum_{k=1}^{K}\theta^{k} \leq 1$ for each platoon. The PL's groupcast rate is limited by the vehicle with the worst reception channel quality in the platoon. The effective groupcast rate of $i^{th}$ PL is
\begin{equation}
    \mathrm{C}_{\mathrm{i}}^{k}=\min _{j \in \mathrm{M}}\left\{\log _{2}\left(1+\gamma_{0, j}^{k}\right)\right\}
    \end{equation}
where $\gamma_{0, j}^{k}$ represents the SINR between PL and his PM $j$. Because the signal attenuation is mainly caused by wireless propagation loss, it can be considered that the tail vehicle has the worst reception channel quality \cite{b4}. Then the above formula can be abbreviated as $\mathrm{C}_{i}^{k}=\log _{2}\left(1+\gamma_{0, M}^{k}\right)$

\subsection{Problem Formulation}
Due to the critical role of V2V communication in the platoon safe driving process, the platoon has strict requirements on the reliability and delay of communications. Since the safety of the platoon requires the message of PL, the groupcast quality in the platoon should be guaranteed. Our optimization goal is to maximize the reliability of platoon groupcast, which can be expressed as follows
\begin{equation}
    \begin{array}{c}
    \max_{\theta_{i}^{k}, P_i} \operatorname{Pr}\left\{\sum_{t=1}^{T} \sum_{i=1}^{N} \sum_{k=1}^{K} \theta_{i}^{k} C_{i}^{k} \Delta_{T} \geq B_{pkt}\right\} \\
    \\
    \text {s.t.}\left\{\begin{array}{l}
    \sum_{k=1}^{K} \theta_{i}^{k} \leq 1, \theta_{i}^{k} \in\{0,1\} \\
    \\
    0\leq P_i \leq P_{max}
    \end{array}\right.
    \end{array}
    \end{equation}
where $B_{pkt}$ is the payload size, $T$ is the payload generation period, and $\Delta_{T}$ represents unit time slot. Limit $1$ indicates the maximum number of subchannel that can be selected in each groupcast. Limit $2$ represents a numerical limit on the transmit power of PL. Due to the high computational complexity and the random evolution of channel conditions, it is difficult to obtain a globally optimal solution in practical applications. Therefore, we propose an optimization algorithm based on VDN to solve this problem.

\section{Multi-agent Deep Reinforcement Learning based Resource Allocation}
We assume that each PL only obtains the channel state information (CSI) of its platoon before taking action \cite{b9}. Each PL uses a distributed resource selection strategy to select resources independently. In this section, we first model this optimization problem as Markov games and then propose a distributed resource allocation algorithm based on VDN to solve the resource allocation problem.

\subsection{Multi-agent Environment Modeling}
In reinforcement learning (RL), an agent learns strategies through interaction with the environment to achieve the goal of maximizing returns. Mathematically, reinforcement learning problems can be modeled as Markov Decision Processes (MDP). In particular, the above process can be described as a tuple $(\mathcal{S}, \mathcal{A}, \mathcal{R}, \mathcal{P})$. The reward function $r = R (s_t, a_t|s_{t+1})$ can be obtained according to the current state $s_t \in \mathcal{S}$, the taken action $a_t \in \mathcal{A}$, and the state $s_{t+1} \in \mathcal{S}$ corresponding to the next moment. $\mathcal{P}$ of the MDP indicates the probability of the next state is $s_{t+1}$ when the action $a_t$ is taken under the state $s_t$, which can be expressed as $p(s_{t+1} | s_t, a_t)$. Unfortunately, traditional reinforcement learning methods such as Q-learning and Policy Gradient-based on MDP are not applicable in distributed resource allocation. The major problem is that as the training progresses, each agent independently updates its strategy, which affects the state transition probability of other PLs. from any agent's perspective, the environment becomes unstable, which violates the Markov assumptions required for convergence of RL \cite{b5}.

In order to overcome the convergence problem of single-agent MDP, the environment is modeled as Markov games in this paper. Markov games is an extension of the MDP in the case of multi-agent. It can be expressed as a tuple: $(\mathcal{V}, \mathcal{S}, \mathcal{A}, \mathcal{R}, \mathcal{P})$, where $\mathcal{V}$ is participants in the Markov games, $\mathcal{S}$ is the state space, $\mathcal{A}$ is the action space of all individuals in the environment, and $\mathcal{R}$ is the reward function. In collaborative Markov games, the goal of each agent $i$ is to maximize the overall reward $\mathrm{R}=\sum_{t=0}^{T} \mu_{t} r_{t}$ of the system, where $0 \leq \mu_{t} \leq 1$ is a discount factor. The larger $\mu_{t}$ means more attention to future rewards. $T$ represents the time frame for calculating rewards. At every discrete moment $t$, different agents get different descriptions of the environment $\{o_1, o_2, ..., o_n\}$ based on own partially observation. Note that a single observation cannot represent the entire environment. Agent $i$ chooses action $a_i$ based on strategies $\pi_i$. The joint action behavior of multi-agent can be expressed as $\mathbf{a}(s)=a_{1}(o_1) \times a_{2}(o_2) \times \cdots \times a_{N}(o_N)$, and the corresponding state transition probability $P$ is $p(s_{t+1} | s_t, \mathbf{a})$. In Markov games, each agent needs to consider the strategies of other agents while learning interactively with the environment.

In our solution, the state space, the action space, and the reward function is defined as follows:

\textbf{Observation Space:} At each time slot t, each agent observes local information, which is used to describe the current state of the environment. It consists of internal and external parts. The internal information includes the remaining load $B_{i}^{t}$ of the groupcast payload and the remaining time budget $T_{i}^{t}$. The external information is the interference situation $I_{i}^{t-1}$ of the platoon $i$ in all subchannels at a previous moment, which can be measured by PMs. Therefore, the observed state of agent $i$ at time slot $t$ is
\begin{equation}
    o_{i}^{t}=\left\{I_{i}^{t-1}, B_{i}^{t}, T_{i}^{t}\right\}
    \end{equation}

\begin{figure}[htbp]
    \centerline{\includegraphics[scale=0.4]{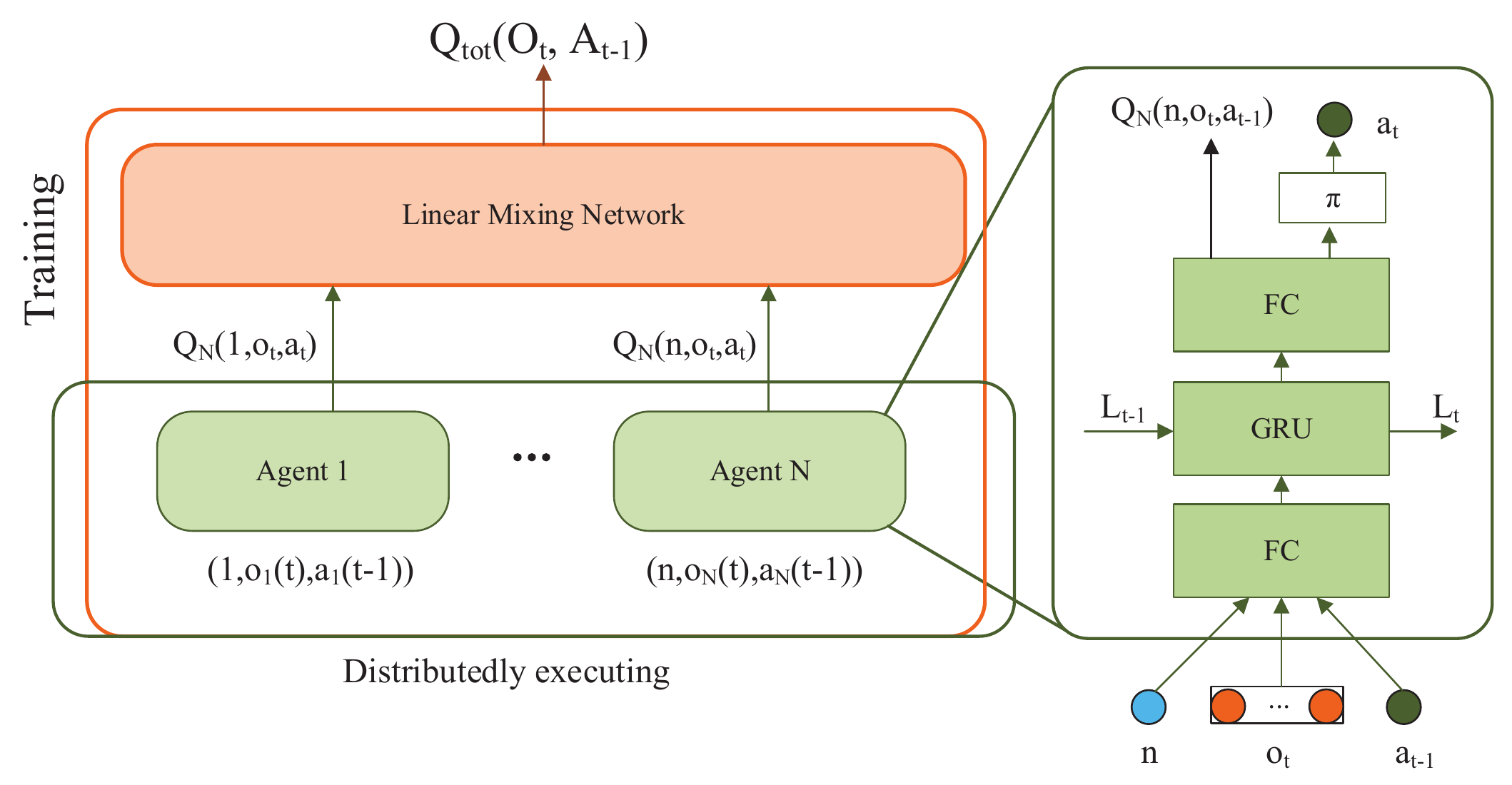}}
    \caption{Architecture of VDN based resource selection algorithm.}
    \label{fig2}
\end{figure}
    
\textbf{Action Space:} Agent $i$ takes action $a_i^t$ at time $t$, which indicates the subchannel and power selected by the agent according to the strategy and the observed state. The dimension of the action space is $K * H$, because there are $K$ subchannels and $H$ discrete power levels.

\textbf{Reward:} Our optimization goal is to maximize the probability of successful transmission of V2V loads. To avoid learning difficulties caused by sparse rewards, we set reward to an effective V2V transmission rate during the agent payload transmission process. We use prior knowledge to design the reward value, that is, a higher transmission rate can complete the transmission of the payload more effectively and accelerate the learning of the agent \cite{b10}. In our tunning experience, we found that the agent will deliberately score points, such as the curve reaching the end. Therefore, after all, payload transmission is completed $B_i^t = 0, \forall i \in N$, we give the system a reward that is positively related to the remaining time budget. So in each time slot, the relevant reward is offered by
\begin{equation}
    R(t)=\left\{\begin{array}{ll}
    \sum_{i=1}^{N}\sum_{k=1}^{K} \theta_{i}^{k} C_{i}^{k}, & \text { if } B_{t}^{i} > 0, \forall i \in N\\
    \\
    \tau T_t, & \text { otherwise }
    \end{array}\right.
    \end{equation}
where $\tau$ is hyperparameters that need to be tuned in experiments, note that all platoons use the same reward.

\subsection{Multi-agent Interference Control Scheme}
Aiming at the problem that distributed resource selection, we studied the problem of maximizing system reward based on Markov games and proposed a VDN-based resource selection algorithm for platoon resource selection. The main assumption in the VDN framework is that the system's joint action-value function can be decomposed into different agent value functions \cite{b11}
\begin{equation}
    Q_{tot}\left(\left(o_{1}, \ldots, o_{N}\right),\left(a_{1}, \ldots, a_{N}\right)\right) \approx \sum_{i=1}^{N} Q_{i}\left(o_{i}, a_{i}\right)
    \end{equation}
$Q_{i}$ is the action value based on the agent's local observation, and $Q_{tot}$ represents the joint action value of the system. In the algorithm learning process, a joint $Q$ network is trained centrally, which is obtained by the aggregation of local $Q$ networks of all agents. Specific network architecture, as shown in Fig.~\ref{fig2}. 

Importantly, to ensure that the global max operations performed on $Q_{tot}$ has the same result as the single max operation performed on $Q_{i}$ of each agent. The monotonicity of the linear hybrid network needs to be guaranteed
\begin{equation}
    \frac{\partial Q_{t o t}}{\partial Q_{i}} \geq 0
    \end{equation}

During the training process, the joint action-value function is used for back-propagating by different individual deep neural networks, which can promote agents to cooperate to achieve the overall optimal \cite{b11}. After the training is completed, each agent gets a $Q$ network based on its local observations, used for decentralized execution.

To tackle the problem of partially observable in Deep Q-learning (DQN), we input the current local observations of the agent $i$ and the actions $a_{i}^{t-1}$ of the previous moment into the DQN. Furthermore, we added Gate Recurrent Unit (GRU)  after fully connected layers (FC) \cite{b12}. GRU is a special kind of RNN that can solve problems such as long-term memory. Hence, agents can integrate information from previous moments to make up for the shortcomings of current local observations. As shown in Fig.2, we use the output of FC and the output $L_{t-1}$ of GRU as input to the GRU. Then GRU generates the output $L_{t}$ at the current time. Such feedback can make better use of local observations to approximate the actual $Q$ value.

As the number of agents increases, the number of learnable parameters also increases. Then the computational complexity, memory storage, and processing time during training become uncontrollable. Inspired by \cite{b13}, we let each agent share network parameters during the learning process since they are isomorphic. By adding agent identification at the input, a DQN network can be used for multi-agent. While avoiding lazy agents, it can reduce processing delay and accelerate network convergence. However, it should be noted that each agent's learning strategy is based on their history. Role information can be provided to the network in the form of 1-hot encoding.

Nevertheless, as the number of agents increases, the size of the vector becomes more extensive, and non-zero values in 1-hot become considerably sparse. Under these circumstances, directly coding through 1-hot will still increase the burden of training. So when the number of agents exceeds a specific number, we encode the role information of the agent to reduce space consumption. Then, we connect the encoded id to the previous action and the current observation state of the corresponding agent and input it into VDN.

\subsection{Training Implementation}

\begin{algorithm}[htb] 
    \caption{The VDN-based Solution} 
    \label{alg:Framwork} 
    \begin{algorithmic}[1] 
    \REQUIRE  
    DQN network structure, hybrid network structure.
    \ENSURE  
    The weights of DQN network.
    
\STATE Initialize multi-platoon simulation environment, train network parameters $\omega$, target network $\omega^{'} \leftarrow \omega$ and initialize the replay memory $\mathcal{D}$;
\FOR{each $episode$} 
\STATE Update simulation environment and exploration rate $\varepsilon$;
\FOR{each $step$} 
\STATE Receive observation state $s_{t}=\left\{o_1, o_2, \dots, o_N \right\}$;
\FOR{$agent \in [1,N]$}
\STATE With probability $\varepsilon$ pick random $a_i^t$;
\STATE else $a_{i}^{t}=\arg \max _{a \in \mathcal{A}} Q\left(i, o_{i}^{t}, a_{i}^{t-1}, L_{i}^{t-1}; \omega_{i}\right)$;
\ENDFOR
\STATE Perform joint actions $\mathbf{a}(s)$ to get reward $r_t$ and new state $s_{t+1}$;
\STATE Store $(s_t, a_t, r_t, s_{t+1})$ in the replay memory $\mathcal{D}$ in sequence, then ($s_i \leftarrow s_i^{'}$);
\ENDFOR
\STATE Randomly select period data $(s_t, a_t, r_t, s_{t+1})$ from $\mathcal{D}$;
\STATE Train the VDN network through equation (10);
\STATE Update target Q-network parameters ($\omega^{'}\leftarrow \omega$);
\ENDFOR
\end{algorithmic}
\end{algorithm}

Similar to \cite{b10}, our scheme uses centralized training and distributed execution. In particular, each PL independently runs its DQN when distributed implementation. 

To ensure a balance between exploration and development, we adopt the $\varepsilon$-greedy exploration. The probability of $1-\varepsilon$ that PL adopts the optimal action output by VDN and the probability of $\varepsilon$ is that PL chooses the action randomly to realize the exploration of the environment \cite{b14}. In order to make sufficient use of the learning results of VDN, we linked the exploration rate $\varepsilon$ to the training episodes $\varepsilon = 1 - \Delta\varepsilon * episode$, if $\varepsilon > \varepsilon_{min}$.

When corresponding joint operation $a_t = \{a_1^t, ..., a_N^t\}$ is executed, the observation value of each agent is changed to $\{o_1^{t+1}, \dots, o_N^{t+1}\}$. Moreover, each agent received a unified reward of the environment. We deposit transformation matrices $<S, A, R, S'>$ of each agent in each step during each episode into replay memory $\mathcal{D}$ sequentially. Different from traditional DQN, to take advantage of the temporal properties of GRU, we randomly extract multiple full-episode data from $\mathcal{D}$ for training during strategy learning. Similar to DQN, we designed two neural networks to cut the correlation and improve the convergence. In particular, each agent calculates the target Q-value through the Q-target network and reward

\begin{equation}
    y_i=r_{i}+ \eta Q_{target}(i, o_{i}^{t+1}, a_{i}^{t}|\omega_i^{'})
\end{equation}
where $r_{i}$ is the corresponding reward, and $\eta$ is a discount factor. Then using the difference of overall target Q-value and estimated Q-value to update the training network
\begin{equation}
    \mathcal{L}\left(\omega\right)=\mathbb{E}\left[\left(\sum_{i=1}^{N}\left(Q_{i}\left(o_{i}^{t}, a_{i}^{t}|\omega_i\right)-y_i\right)\right)^{2}\right]
    \end{equation}

\begin{table}[htbp]
    \caption{Simulation Parameters}
    \begin{center}
    \begin{tabular}{|c|c|}
    \hline
    \textbf{Parameter}&\textbf{Value} \\
    \hline
    Number of V2V subchannel& 2\\
    \hline
    subchannel bandwidth& 180kHz\\
    \hline
    Carrier Frequency& 5.9 GHz\\
    \hline
    Number of platoon& 4\\
    \hline
    V2V payload size& [1,2,...] * 1200 bytes\\
    \hline
    V2V transmit power& [23, 15, 10, -114] dBm\\
    \hline
    V2V payload generation period T& 100ms\\
    \hline
    Large-scale fading update& WINNER+B1 every 100 ms\\
    \hline
    Small-scale fading update& Rayleigh fading every 1 ms\\
    \hline
    \end{tabular}
    \label{tab1}
    \end{center}
    \end{table}

Every specific training episode, the training network $\omega$ is used to update the parameters of the target network $\omega^{'}$: $\omega^{'}\leftarrow \omega$. Algorithm 1 describes in detail the multi-agent resource selection algorithm based on VDN in the multi-platoon cooperative interference control scenario.

\section{Simulation}
Our multi-platoon Cooperative interference control simulation platform is built according to the highway scenario of 3GPP TR 37.885 \cite{b15}. The distance between the head and tail vehicles in the platoon is 75m. Platoons are evenly distributed on two adjacent roads. PL independently selects subchannels and transmission power based on their observations and strategies to complete the groupcast within the platoon. More detailed simulation parameter settings are shown in Table~\ref{tab1}. The agent's minimum exploration rate is 0.03. The number of neurons in the hidden layer GRU is 64. Other fully connected layers use the rectified linear unit (ReLU) as the activation function. The learning rate is $0.001$. we randomly sample 32 episodes of data each time for training. All simulations are implemented on PyTorch $1.4$ and Python $3.6$. To verify the effectiveness of the proposed algorithm, we will compare two baseline scenarios. 1) MARL: Each agent has a fingerprint-based deep Q-network, which is used to deal with the instability of the environment \cite{b8}; 2) Random: Each agent randomly selects the subchannels and transmit power, at each time step.

\begin{figure}[htbp]
    \centerline{\includegraphics[scale=0.6]{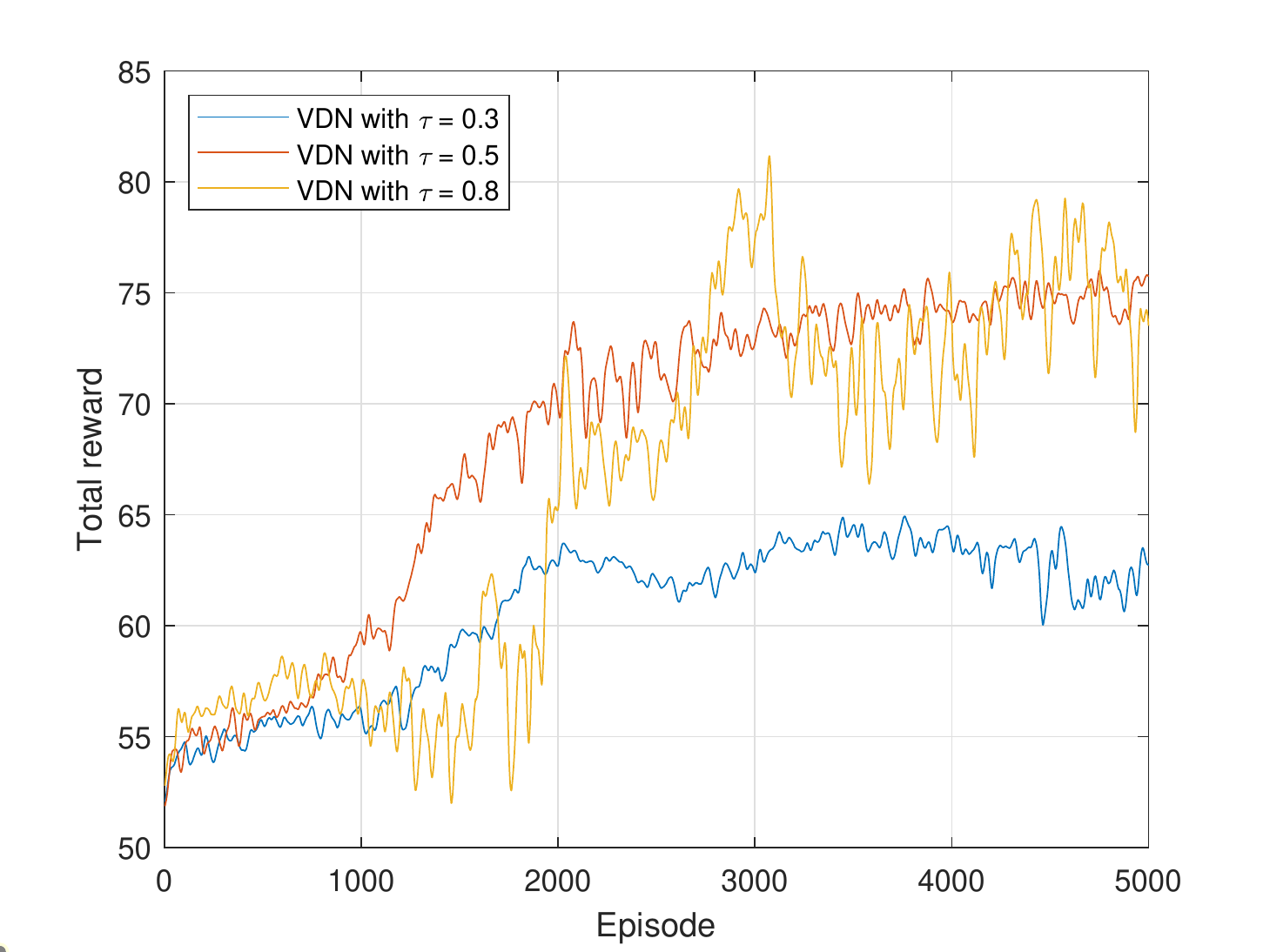}}
    \caption{Total reward of the algorithm during the training process with different $\tau$. The platoon payload size $B = 6000$ bytes.}
    \label{fig3}
\end{figure}

Fig.~\ref{fig3} shows the convergence of the proposed algorithm. As the training progresses, the exploration rate gradually decreases, and the agent learns strategies from the environment. The total reward increases with the training cycle, and eventually stabilizes within a specific range, which demonstrates the effectiveness of the proposed algorithm in solving Markov games. Further, we compare the role of the hyperparameter $\tau$. When $\tau = 0.3$, the algorithm can converge, but its reward remains at a lower level. This is because the reward for completing the transfer is not enough to motivate the agent to complete the payload transfer. And the agent may be satisfied with the reward of the transfer process, resulting in the transmission failure. But when $\tau = 0.8$, the training reward fluctuates severely, which is because the reward for completing the transmission is too large compared to the previous incentive reward—cause network training shock. When $\tau = 0.5$, the agent can get a suitable completion reward, which can not only ensure the convergence of the network but also motivate the agent to complete the payload transmission as soon as possible. Therefore, in the application of the algorithm, we need to set the coefficient $\tau$ reasonably to make full use of the algorithm performance.
    
\begin{figure}[htbp]
    \centerline{\includegraphics[scale=0.6]{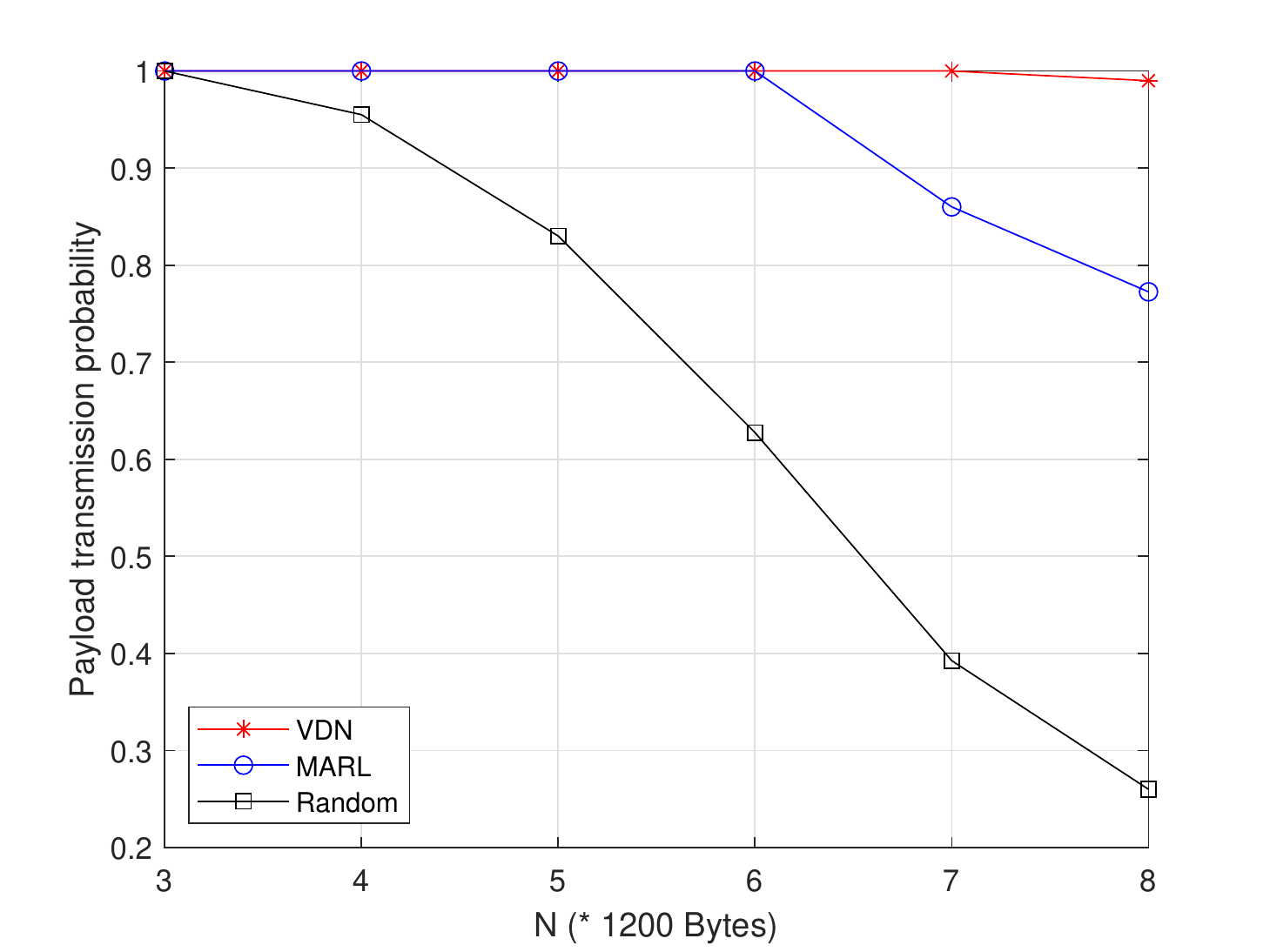}}
    \caption{Platoons payload delivery probability with varying payload sizes.}
    \label{fig4}
\end{figure}
\begin{figure}[htbp]
    \centerline{\includegraphics[scale=0.6]{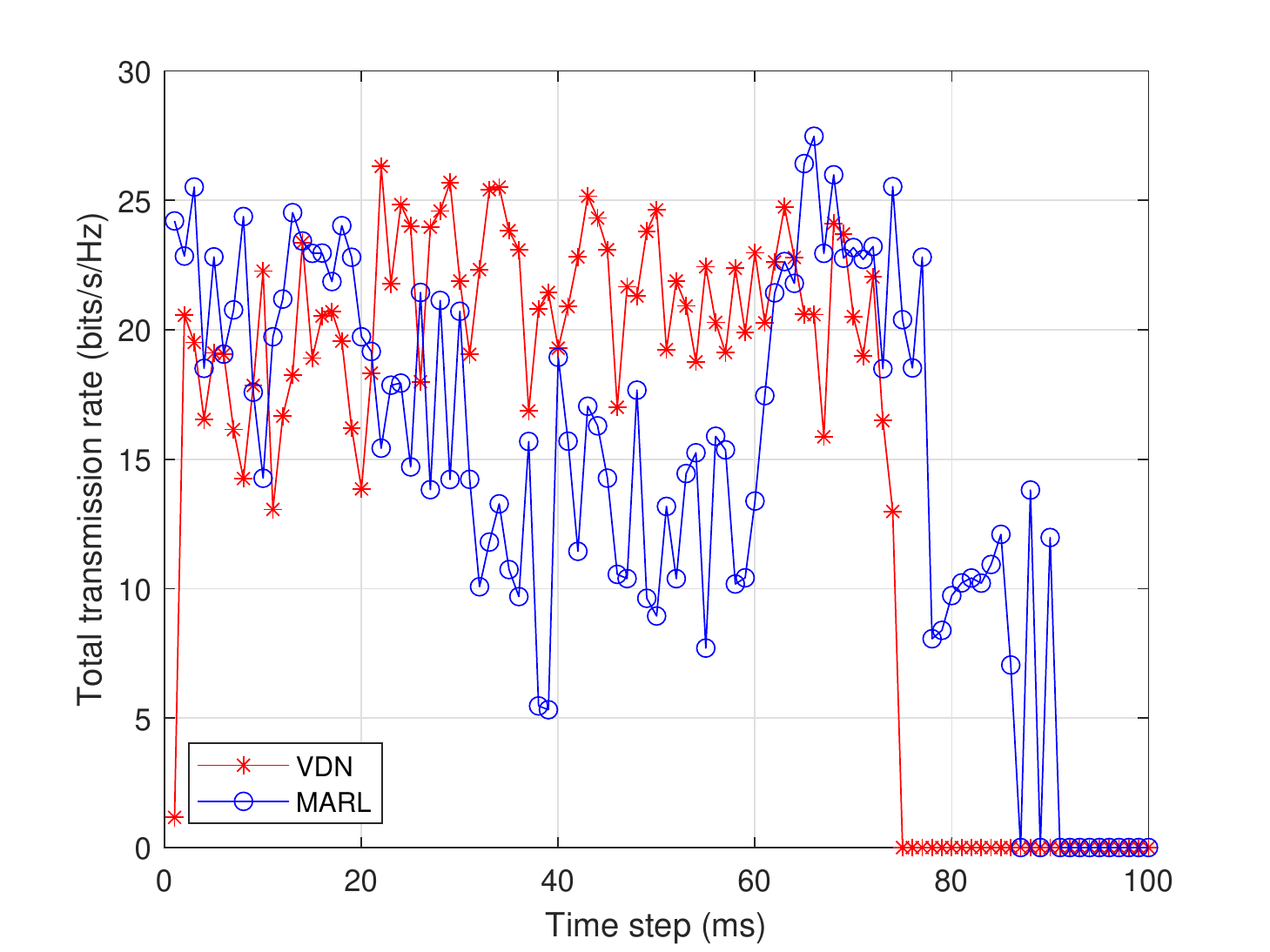}}
    \caption{Total platoons transmission rates of the different schemes within the same episode. The platoon payload size $B=8400$ bytes.}
    \label{fig5}
\end{figure}

The relationship between the success rate of the platoon groupcast and the payload size of the platoon groupcast is demonstrated in Fig.~\ref{fig4}. With the increase of the payload, the transmission success rates of different algorithms have declined to vary degrees. Among them, the Random algorithm has the worst performance, and transmission failure occurs when $N = 4$, which also illustrates the necessity of a resource allocation algorithm in distributed scenarios. For the MARL algorithm, a large number of transmission failures also occur when $N = 7$, but it has been improved relative to Random. For the proposed algorithm, a small number of failures did not happen until $N = 8$. Further observing the characteristics of the different curves, it can be found that the slope of the proposed algorithm is the smoothest as $N$ increases. This shows that even if there is a large amount of payload that the algorithm cannot handle, the proposed algorithm can still meet the system performance requirements to the greatest extent.

To further explore the reasons for the different performance of different algorithms in Fig.~\ref{fig4}, we randomly select a transmission episode. Then in Fig.~\ref{fig5}, the total rate of system transmission for each step in this transmission episode is plotted. The results show that the proposed algorithm has excellent performance. In the MARL scheme, although there is a high transmission rate at the beginning, it cannot be maintained. It fluctuates constantly, and even a zero transmission rate situation occurs during transmission. There are two main reasons for this phenomenon: the agent cannot timely respond to changes in the environment due to changes in the transmission channel caused by vehicle movements. The other reason is that MARL's distributed ability to coordinate different agents resource selection strategies is limited, resulting in selected channels resource conflict or non-optimal sending power. As a result, the transmission rate fluctuates severely in MARL. In the proposed scheme, although the system transfer rate is low at the beginning. However, as time progresses, the system transmission rate is stable at a high level. Note that, the reason why our solution has a total transmission rate of $0$ after time step $74$ is that PLs have transmitted his payload, not because of collision. The results show that with the proposed algorithm, even if there is no signaling interaction between different agents, the resource selection of different agents can achieve distributed coordination through interaction with the environment. It avoids the occurrence of resource conflicts, which improves the efficiency of the system.

\section{Conclusion}
This paper studies the groupcast channel resource selection problem in a multi-platoon scenario. We model the problem as Markov games, then propose a VDN based multi-agent distributed resource allocation scheme. The proposed solution does not require signaling interaction between agents during the execution phase and can achieve cooperation between agents to achieve global optimization. At the same time, the scheme can decrease the training delay by sharing the neural network parameters between agents. Simulation results show that the solution has good convergence and can effectively improve the probability of successful transmission of the platoon. We believe that the algorithm proposed in this paper can be implemented in other similar multi-agent scenarios.

\end{document}